\documentclass[letter]{aa} 
\pdfoutput=1

\usepackage{graphicx}
\usepackage{xtab,booktabs}
\begin{document}
   \title{Physical and orbital properties of $\beta$ Pictoris b}

\titlerunning{Physical and orbital properties of $\beta$ Pictoris b}
\authorrunning{Bonnefoy et al.}

\author{M. Bonnefoy \inst{1}, G.-D. Marleau\inst{2}, R. Galicher\inst{3}, H. Beust \inst{1}, A.-M. Lagrange\inst{1}, J.-L. Baudino\inst{3}, G. Chauvin \inst{1}, S. Borgniet\inst{1}, N. Meunier\inst{1}, J. Rameau\inst{1}, A. Boccaletti\inst{3}, A. Cumming\inst{4}, C. Helling\inst{5}, D. Homeier\inst{6}, F. Allard\inst{6}, P. Delorme\inst{1}}

\institute{
$^{1}$ Universit\'{e} Grenoble Alpes, IPAG, F-38000 Grenoble, France. CNRS, IPAG, F-38000 Grenoble, France \\
$^{2}$ Max Planck Institute for Astronomy, K\"{o}nigstuhl 17, D-69117 Heidelberg, Germany \\
$^{3}$ LESIA, CNRS, Observatoire de Paris, Univ. Paris Diderot, UPMC, 5 place Jules Janssen, 92190 Meudon, France \\
$^{4}$ Department of Physics, McGill University, 3600 rue University, Montr\'{e}al, Qu\'{e}bec H3A 2T8, Canada\\
$^{5}$ SUPA, School of Physics and Astronomy, University of St Andrews, St Andrews, KY16 9SS, UK  \\
$^{6}$ CRAL, UMR 5574, CNRS, Universit\'{e} de Lyon, \'{E}cole Normale Sup\'{e}rieure de Lyon, 46 All\'{e}e d'Italie, F-69364 Lyon Cedex 07, France\\
}
   \date{Received 22/04/2014; accepted 13/07/2014}

\abstract{The intermediate-mass star \object{$\beta$ Pictoris} is known to be surrounded by  a structured edge-on debris disk within which a gas giant planet was discovered orbiting at  8-10 AU. The physical properties of \object{$\beta$ Pic b} were previously inferred from broad- and narrow-band 0.9-4.8 $\mu$m photometry. We used  commissioning data of the \textit{Gemini Planet Imager} (GPI) to obtain new astrometry and a low-resolution (R$\sim$35-39) J-band (1.12-1.35$\mu$m) spectrum of the planet. We find that the planet has passed the quadrature. We constrain its semi-major axis to $\leq$10 AU (90\% prob.) with a  peak at $8.9^{+0.4}_{-0.6}$ AU. The joint fit of the planet astrometry and the most recent radial velocity measurements of the star yields a planet dynamical mass lower than 20 $\mathrm{M_{Jup}}$ ($\geq$96\% prob.). The extracted spectrum of $\beta$ Pic b is  similar to those of young $\mathrm{L1_{-1.5}^{+1}}$ dwarfs. We used the spectral type estimate to revise the planet luminosity to $\mathrm{log(L/L_{\odot})=-3.90\pm0.07}$. The 0.9-4.8$\mu$m photometry and spectrum are reproduced  for $\mathrm{T_{\rm eff}=1650 \pm 150}$ K and a log g $\mathrm{\leq 4.7}$ dex by 12 grids of PHOENIX-based and LESIA atmospheric models. For the most recent system age estimate ($21\pm4$ Myr), the bolometric  luminosity and the constraints on the dynamical mass of $\beta$ Pic b are only reproduced by warm- and hot-start tracks with initial entropies $S_{i}>10.5~ k_{B}/baryon$. These initial conditions may result from an inefficient accretion shock and/or a planetesimal density at formation higher than in the classical core-accretion model. Considering a younger age for the system or  a conservative formation time for $\beta$ Pic b does not change these conclusions.}


\maketitle
%

%
\section{Introduction}
A  candidate giant planet  was identified  in 2003 high-resolution imaging data at a projected separation of 9 AU in the disk of the intermediate-mass star $\beta$ Pictoris \citep{2009A&A...493L..21L}. Follow-up images of the system  with various instruments \citep{2010Sci...329...57L, 2013A&A...551L..14B, 2014ApJ...786...32M} from 0.98 $\mu$m to 4.8 $\mu$m enabled us to confirm that $\beta$ Pic b is bound to the star and has a hot ($T_{\rm eff}\sim1700$ K) and dusty atmosphere \citep[][and references therein]{2013A&A...555A.107B, 2013ApJ...776...15C, 2014ApJ...786...32M}.  The monitoring of the planet's orbital motion restrained the semi-major axis (s.m.a.) estimates to 8-10 AU  \citep{2012A&A...542A..41C, 2013A&A...559L..12A}. Combined with radial velocity (RV) measurements \citep{2012A&A...542A..18L,2013A&A...559A..83L}, the  s.m.a. $\leq 10$ AU (80\% prob.) set an upper limit of 15.5 M$\mathrm{_{Jup}}$ on the mass of $\beta$ Pic b \textit{for the case of a circular orbit}. 

The dynamical mass constraints, the $T_{\rm eff}$ and luminosity of the planet, and the system age provide a so far unique test of evolutionary models predictions.  ``Hot-start" models predict $\beta$ Pic b to be a  8 to 12.6 $\mathrm{M_{Jup}}$ planet. "Cold-start" models assume that the gravitational potential energy of the infalling gas at formation is fully radiated away at a supercritical accretion shock. These tracks cannot reproduce the measured photometry of $\beta$ Pic b for planet masses below 15.5 M$\mathrm{_{Jup}}$. ``Warm-start" models  \citep[][hereafter SB12 and MC14]{2012ApJ...745..174S, 2014MNRAS.437.1378M} explore the sensitivity of the mass prediction to the initial conditions, parametrized by the choice of an initial entropy ($S_{\rm i}$). \cite{2013A&A...555A.107B} and \cite{2014MNRAS.437.1378M}, demonstrated that the properties of $\beta$ Pic b can only be reproduced for $S_{\rm i}\geq9.3\:\mathrm{k_{B}}$/baryon, i.e. initial conditions intermediate between cold and hot-start cases. But these predictions relied 1)  on a system age of  $12^{+8}_{-4}$ Myr \citep[][]{2001ApJ...562L..87Z} and 2) on the hypothesis of a non-eccentric orbit for the planet. Since then, \cite{2014MNRAS.438L..11B} reported a lithium depletion  age of $21\pm4$ Myr for the $\beta$ Pictoris moving group. 

In this letter, we present new astrometry and the first J-band spectrum of  $\beta$ Pic b extracted from commissioning data of the Gemini Planet Imager \citep[][]{2014arXiv1403.7520M}  instrument (Section \ref{Section:1}). We use these data in Section \ref{section:Orbit} and up-to-date radial-velocity measurements on the star to refine the constraints on the orbital elements, on the dynamical mass (Section \ref{section:Orbit}), the physical properties, and ultimately, the formation conditions of the planet (Section \ref{section:phys}).
\section{Observations and data reduction}
\label{Section:1}
The source was observed with GPI on Dec. 10, 2013.  The observations combined integral field spectroscopy with  apodized Lyot coronagraphy (diameter=184 mas) and angular differential imaging \cite[ADI,][]{2006ApJ...641..556M}. The data set is composed of 19 spectral cubes consisting of 37 spectral channels each. They cover the J band (1.12-1.35 $\mu$m) at a low resolving power (R=35-39). Data were obtained under good conditions ($\langle\tau_{0}\rangle=14.5$ ms, DIMM seeing=0.7"), low average airmass (1.08), and covered a field rotation of $19.8^{\circ}$.

We used the spectral cubes provided by the GPI pipeline\footnote{http://www.gemini.edu/sciops/instruments/gpi/public-data}. To further process the data, we first registered each slice of the cubes
using the barycenter of the four satellite spots  \citep[attenuated replica of the central star PSF induced by a grid placed in the pupil plane,][]{2006ApJ...647..612M}. We minimized the speckle noise in each slice using the IPAG and LESIA ADI pipelines (whithout spectral differential imaging to minimize biases on the extracted photometry).  The IPAG pipeline used the cADI, sADI, and LOCI methods \citep[see][and references therein]{2012A&A...542A..41C}.  The LESIA pipeline relied on the TLOCI algorithm \citep{2014IAUS..299...48M}.

To estimate the planet photometry and astrometry in each
spectral channel, we assessed biases induced by our algorithms by
injecting fake point-sources into the data cubes built from
the average of the four unsaturated spots over the spectral and time sequence \citep{2014arXiv1404.4635G} before applying ADI speckle-suppression
techniques \citep[][]{2011A&A...528L..15B}.  We used the GPI spot-to-central-star flux-ratio that was calibrated in laboratory (9.36 mag at J band) to obtain the planet-to-star contrast in each spectral channel. We multiplied these contrasts by a template spectrum of $\beta$ Pic A to retrieve the planet spectrum. The template was built by taking the mean of A5V and A7V
star spectra from the \cite{1998PASP..110..863P} library \citep[see][]{2014ApJ...786...32M}. We find a synthetic photometry of $J_{2MASS}=14.1\pm0.3$ mag for the planet consistent with the value reported in \cite{2013A&A...555A.107B}.  The photometric error is given by the quadratic sum of the uncertainty on
the spot-to-star contrast ($0.15$ mag; courtesy of the GPI consortium), on the planet flux measurement ($0.06$ mag) and the variation of the spot flux over the full sequence ($0.1$ mag). The uncertainty on the planet flux measurement and the variation of the spot flux were estimated as in \cite{2014arXiv1404.4635G}. The astrometry is reported in Table~\ref{tab:results}. The associated error is the quadratic sum of 
uncertainties on the centroiding accuracy of individual slices (0.3 pixel), the plate scale (0.02 pixel), the planet template fit (0.1
pixel at J), and the North position angle (1 deg; see the GPI instrument page$^{1}$). 

\begin{table}[ht]
\begin{minipage}{\columnwidth}
\caption{Astrometry for $\beta$ Pic b}
\label{tab:results}
\centering
\renewcommand{\footnoterule}{}  
\begin{tabular}{cccc}
\hline\hline       
Platescale  &  True North     & Separation  &   PA  \\
(mas/pixel)  &  (deg)   &   (mas)   &  ($^{\circ}$)  \\
\hline
 $14.3\pm0.3$ & $0.0\pm1.0$  &  $430\pm10$  &  $211.6\pm1.3$  \\
\hline                  
\end{tabular}
\end{minipage}
\end{table}

\section{Orbit and dynamical mass of $\beta$ Pic b}
\label{section:Orbit}

\begin{figure}[htbp]
\centering
\includegraphics[width=7.5cm]{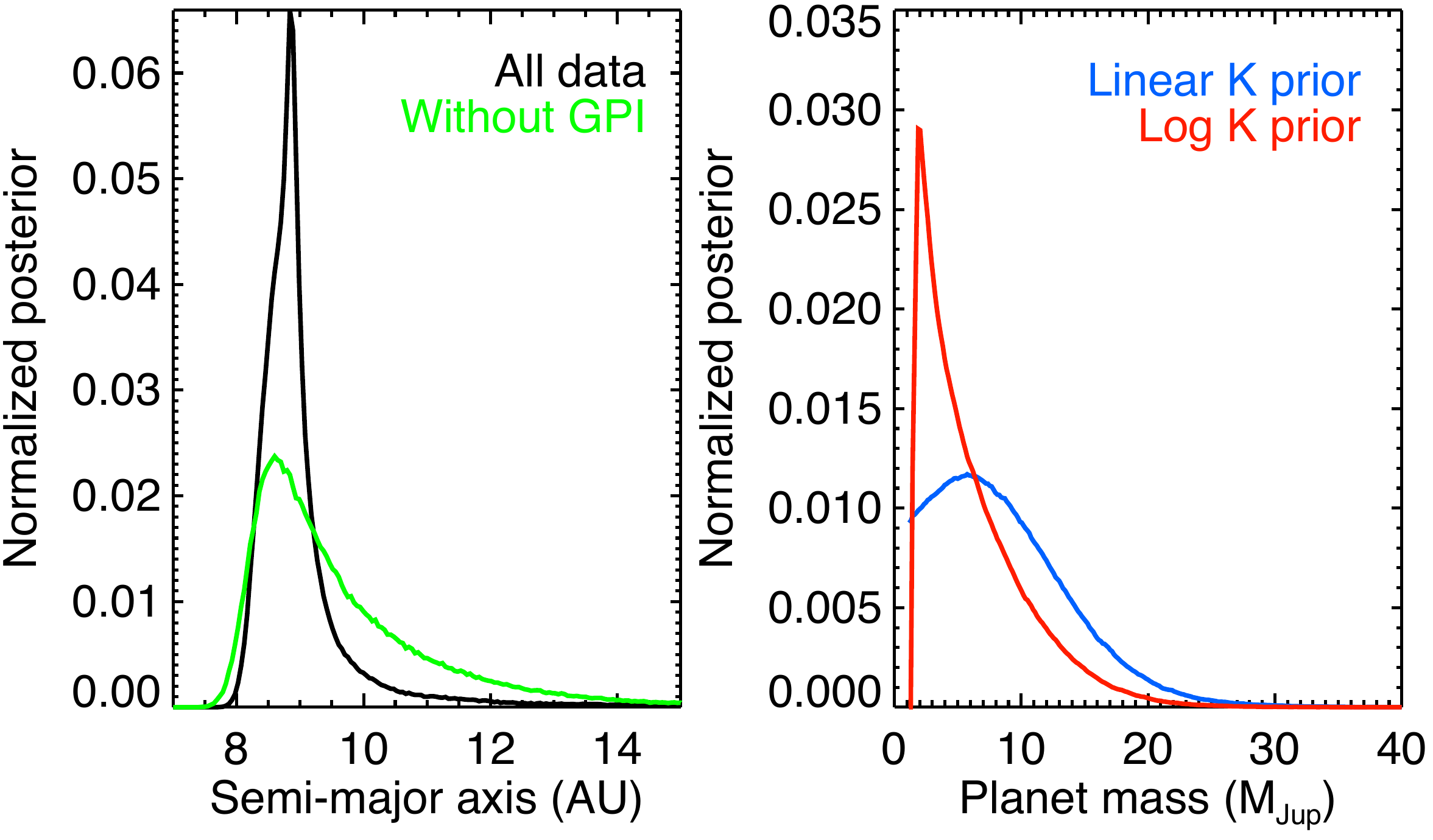}
\caption{Left: MCMC distribution for the semi-major axis of $\beta$ Pic b, with (black curve) and without (green curve) the new GPI data. Right: Dynamical mass distribution of $\beta$ Pic b inferred from the MCMC fit of the combined planet astrometry and RV measurements of the star. The two priors considered here give two different distributions (see Appendix \ref{Appendix:Appmass}).}
\label{Fig:Fig1}
\end{figure}

We combined the GPI relative astrometry of $\beta$ Pic b (Tab \ref{tab:results}) with previous NaCo measurements \citep{2012A&A...542A..41C, 2013A&A...555A.107B, 2013A&A...559L..12A}  to refine the orbital solutions of the object based solely on the astrometry. We used the Markov-chain Monte Carlo (MCMC) Bayesian analysis
technique described in \cite{2012A&A...542A..41C} to derive the
probablilistic distribution of orbital solutions. The new GPI astrometric measurements
confirm  that $\beta$ Pic b has now
passed the quadrature. The semi-major axis distribution is now greatly improved and exhibits a clear peak at
$8.9^{+0.4}_{-0.6}$~AU (Fig. \ref{Fig:Fig1}). The probability distributions of other orbital parameters
remain consistent with the previous estimations of
\cite{2012A&A...542A..41C} and \cite{2013A&A...559L..12A}.  The probability that
$\beta$ Pic~b actually transits along the line of sight is $2\%$. If this is the case, the next
transiting event is expected for mid-2017. These conclusions are consistent with the analysis inferred from GPI H band (1.65 $\mu$m) data of the system obtained on Nov. 18, 2013 \citep{2014arXiv1403.7520M}.


We tried to constrain the mass of $\beta$ Pic b using both the planet  astrometry and an up-to-date compilation of RV measurements (Borgniet et al., in prep) of the system gathered since 2003 with the high-precision spectrometer HARPS. In contrast to the upper limits on the mass derived in \cite{2012A&A...542A..18L},  these dynamical mass estimates do not rely on the hypothesis of a circular orbit any more. To do this, we modified our existing MCMC code \citep{2012A&A...542A..41C} to account for both the astrometric and RV data sets in the $\chi^{2}$ computation. This introduces  two additional parameters in the MCMC simulations: the amplitude $K$ of the RV curve, and an offset velocity. The mass of the planet can be derived from the $K$ value and from the other determined orbital parameters for any orbital solution. Because of the large uncertainty on the RV data, the orbit is still mainly constrained by the astrometric data. Conversely, the mass is constrained by the RV data. We assumed a stellar mass of $1.75\pm0.05 M_{\odot}$. The error on the stellar mass appeared to have only marginal influence on the planet dynamical mass. The posterior distribution of the mass is, however, extremely sensitive to the assumed errors on the RV data and on the prior assumed on the amplitude $K$ (see Appendix \ref{Appendix:Appmass} for details). Figure \ref{Fig:Fig1} shows two histograms of posterior mass distribution, each of them corresponding to the use of a different prior on $K$ (linear and logarithmic). In both cases, up to 96\% of the solutions are below 20  $\mathrm{M_{Jup}}$.

\section{Physical properties and initial conditions}
\label{section:phys}

\begin{figure}
\centering
\includegraphics[width=7.7cm]{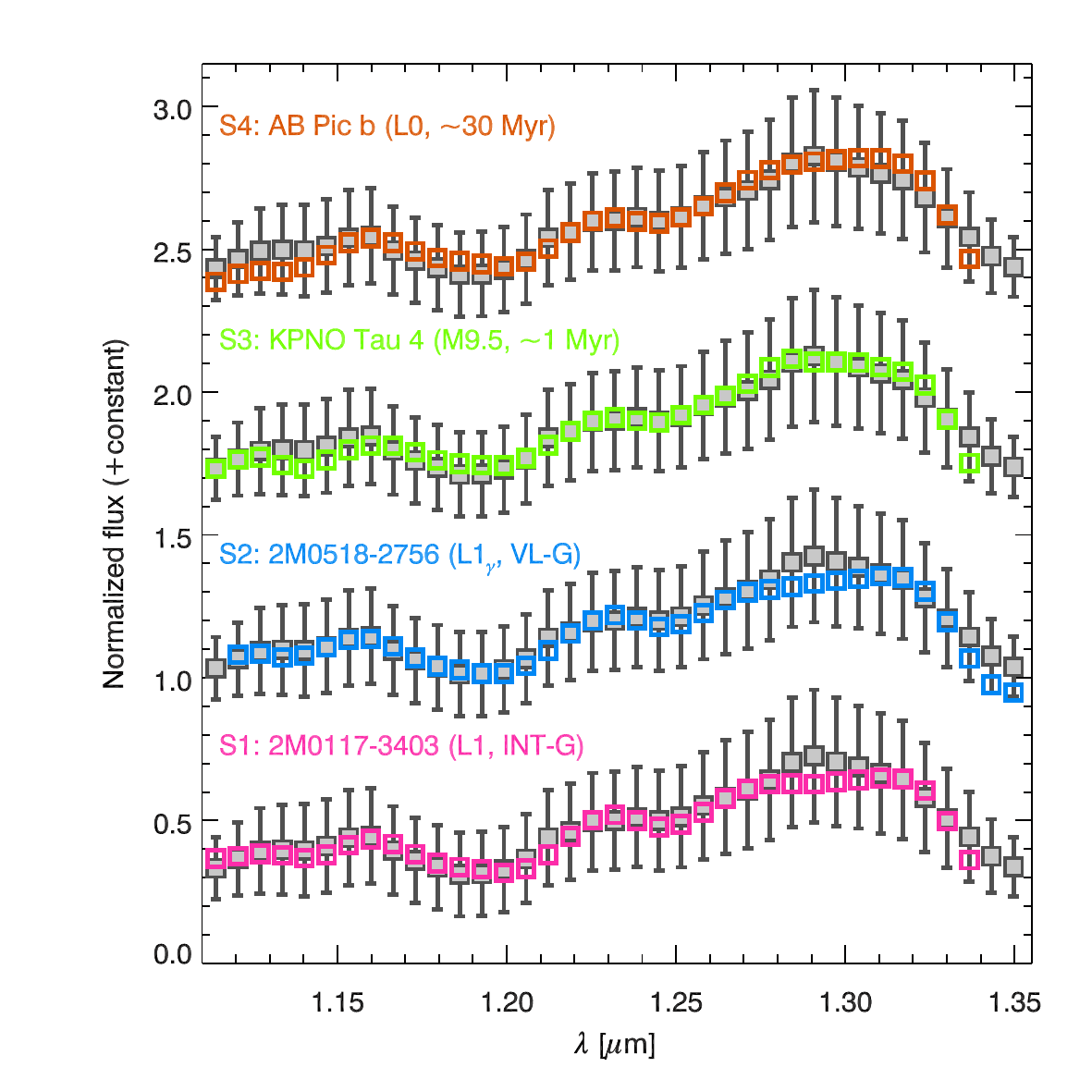}
\caption{Normalized J-band spectrum of $\beta$ Pic b (gray squares) compared with best-fitting spectra (see description in Appendix \ref{Appendix:AppA}) of M9.5-L1 objects from samples 1 (pink squares), 2 (blue squares), 3 (green squares), and 4 (red squares).}
\label{Fig:Fig2}
\end{figure}

The J-band spectrum of $\beta$ Pic b (Fig. \ref{Fig:Fig2}) contains all the feature characteristics of late-M and early-L dwarfs: a marked water-band absorption longward of 1.33 $\mu$m, a rising pseudo-continuum from 1.1 to 1.33 $\mu$m slightly affected from 1.16 to 1.22 $\mu$m by FeH  absorptions. We compared it to four samples of empirical spectra of MLT dwarfs and young planetary mass objects (Appendix \ref{Appendix:AppA}) using a $\chi^{2}$  (Fig. \ref{Fig:Fig2}). In sample 1 (composed of objects of various ages) the spectrum of the candidate AB Dor member (age$\sim$50-150 Myr; J. Faherty, priv. com.)  L1 dwarf \object{2MASSI J0117474-340325} \citep{2008ApJ...681..579B} provides the best fit. Comparisons with the remaining samples confirm that $\beta$ Pic b is an L$1_{-1.5}^{+1}$ dwarf, as previously inferred from the spectral energy distribution (SED) analysis \citep{2013A&A...555A.107B, 2013ApJ...776...15C, 2014ApJ...786...32M}.  

\begin{figure}[htbp]
\centering
\includegraphics[width=8cm]{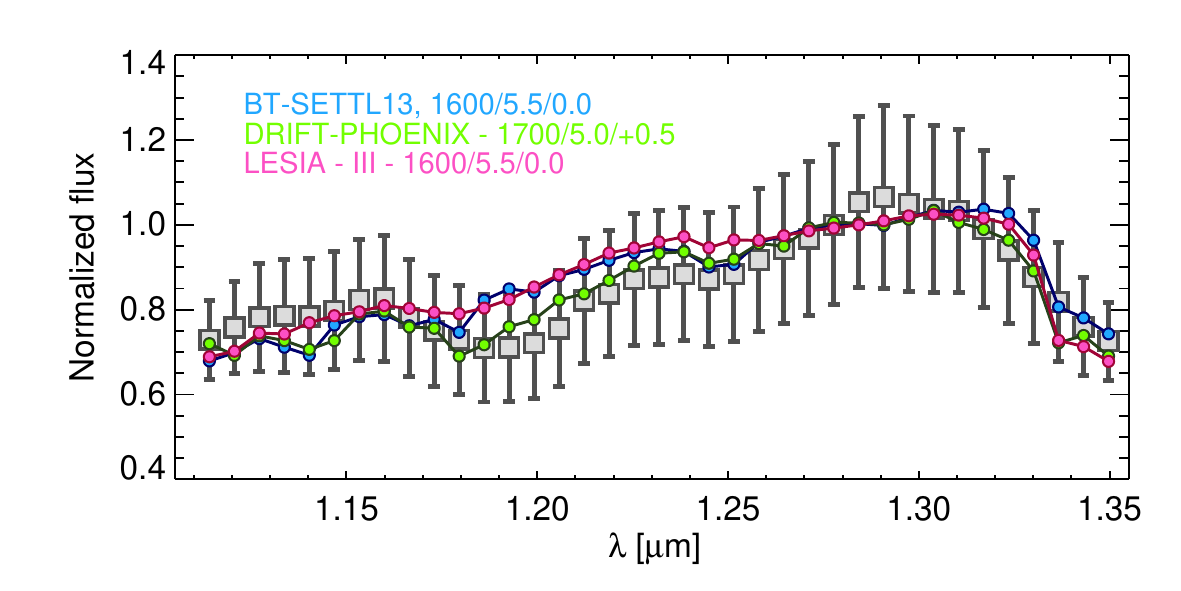} 
\caption{Best-fitting synthetic spectra  from the BT-SETTL13,  DRIFT-PHOENIX, and LESIA grids found from a visual checking. The parameters $T_{\rm eff}$/log g/[M/H] are reported for each model.}
\label{Fig:Fig3}
\end{figure}

We used the bolometric correction of young M9.5-L0 dwarfs (BC$_{K}=3.40\pm0.02$) from \cite{2010ApJ...714L..84T}\footnote{The BC$_{K}$ corresponds to the mean of those measured for KPNO-Tau 4 and \object{2MASS J01415823-4633574}, two objects whose J-band spectra reproduce those of $\beta$ Pic b well.} and the mean and dispersion on the K$_{s}$-band photometry reported in \cite{2014ApJ...786...32M} to find a revised $\mathrm{log(L/L_{\odot})=-3.90\pm0.07}$ for $\beta$ Pic b.   
 We compared the normalized spectrum of $\beta$ Pic b with predictions from the PHOENIX-based  \citep[BT-SETTL10, BT-SETTL13, DRIFT-PHOENIX, described in][]{ 2011A&A...529A..44W, 2013A&A...555A.107B, 2014A&A...564A..55M} and five grids of LESIA atmospheric models \citep[see Appendix \ref{Appendix:AppB} and][]{2014IAUS..299..277B} to derive updated $T_{\rm eff}$ and log g estimates (Table \ref{atmopar}).   A similar analysis derived for the up-to-date SED is reported in Appendix \ref{Appendix:AppC}.  The SED and spectra of  $\beta$ Pic b constrain the $\mathrm{T_{eff}}$ to $\mathrm{1650\pm150}$ K.  The fits are less sensitive to log g and to the metallicity.  Although log g values higher than 4.7 dex cannot be directly excluded from the spectral fitting, these values and the radii derived from $\mathrm{T_{eff}}$ and the luminosity estimates yield masses (see Tables \ref{atmopar} and \ref{atmoparSED}) greater than the dynamical mass constraints (Section \ref{section:Orbit}). The $\chi^{2}$  fit of the J-band spectrum is mostly sensitive to the overal spectral slope and less sensitive to the simultaneous fitting of the water-band absorption longward of 1.3 $\mu$m. Therefore, visual inspection yields similar, but different solutions for the DRIFT-PHOENIX (DP) and LESIA models (Figure \ref{Fig:Fig3}). The  $\mathrm{T_{eff}}$ value is consistent with those reported in \cite{2013A&A...555A.107B}, \cite{2013ApJ...776...15C}, and \cite{2014ApJ...786...32M} using the SED only and/or different atmospheric models. The $\mathrm{T_{eff}}$ is also coherent with those derived for young objects at similar spectral types \citep{2014A&A...562A.127B, 2014A&A...564A..55M}.   The dilution factors needed to adjust the model SED expressed in surface fluxes to the apparent fluxes of the planet correspond to a planetary radius of $1.5\pm0.2$ R$_{\rm Jup}$  \cite[see ][]{2013A&A...555A.107B}. This radius is consistent with those reported in \cite{2013A&A...555A.107B},  \cite{2013ApJ...776...15C}, and with the one derived from the $\mathrm{T_{eff}}$ and luminosity estimates ($\mathrm{1.4^{+0.2}_{-0.1}}$ R$_{\rm Jup}$).

\begin{table}
\begin{minipage}{\columnwidth}
\caption{Best-fitting atmospheric parameters. Solutions leading to semi-empirical masses ( $\mathrm{\mathrm{M_{S.E.}}}$) below 2  $\mathrm{M_{Jup}}$ and above 20 $\mathrm{M_{Jup}}$ are lister in italics (considering uncertainties of $\pm100$ K in $\mathrm{T_{eff}}$,  $\pm0.1$ dex in log g for the LESIA grids,  $\pm0.5$ dex otherwise).}
\label{atmopar}
\centering
\renewcommand{\footnoterule}{}  
\begin{tabular}{lllll}
\hline \hline 
Model		 		  &    $\rm T_{eff}$		&	 $\mathrm{log\:g}$		& $\mathrm{\chi^{2}_{red}}$  & $\mathrm{\mathrm{M_{S.E.}}}$	\\   			
						&				(K)					&		$\mathrm{(cm.s^{-2}}$)	& &	$\mathrm{\mathrm{(M_{Jup})}}$ \\	
\hline
Settl10		&			    1600						&	3.5				&	0.49	& 3$^{+7}_{-2}$  \\
Settl13-M/H=-0.5	&			1500						&	4.5				&	0.38	& 34$^{+86}_{-24}$ \\
Settl13-M/H=0.0	 & 		\textit{1600}						&	\textit{5.5}				&	\textit{0.24}  & 259$^{+643}_{-184}$	\\
Settl13-M/H=+0.5	& 		\textit{1600}						&	\textit{5.0}				&	\textit{0.31} & 82$^{+204}_{-58}$	\\  
DP-M/H=-0.5         &     \textit{1600}            &  \textit{5.5}                   &  \textit{0.20} & 259$^{+643}_{-184}$  \\
DP-M/H=0.0         &  1600                   &  4.5                  &  0.17  & 26$^{+65}_{-19}$  \\
DP-M/H=+0.5         &    1700                &  4.5                  & 0.12  &  20$^{+50}_{-15}$ \\
LESIA - I       &        \textit{2100}         &    \textit{3.6}               &  \textit{2.19} & 1.1$^{+0.3}_{-0.2}$ \\
LESIA - II      &      \textit{1500}               &  \textit{5.5}                 & \textit{1.56}  &  335$^{+138}_{-94}$ \\
LESIA - III     &      \textit{1400}               &    \textit{5.2}               & \textit{0.33} &   221$^{+98}_{-65}$ \\
LESIA - IV    &       \textit{1500}             &   \textit{5.3}               &  \textit{0.25}  &  211$^{+88}_{-59}$ \\
LESIA - V     &       \textit{1500}              &   \textit{5.4}                &  \textit{0.33} &  266$^{+110}_{-75}$ \\
\hline
\end{tabular}
\end{minipage}
\end{table}

The  $T_{\rm eff}$ and luminosity  of $\beta$ Pic b match those of ``hot-start'' evolutionary models  \citep{2000ApJ...542..464C, 2003A&A...402..701B} at an age of $21\pm4$ Myr for $\mathrm{M=11.5\pm0.8\:M_{Jup}}$ and  $\mathrm{M=11.2\pm0.3\:M_{Jup}}$, respectively. This agrees with the mass constraints of Section \ref{section:Orbit}.

\begin{figure}[htbp]
\centering
\includegraphics[width=7.7cm]{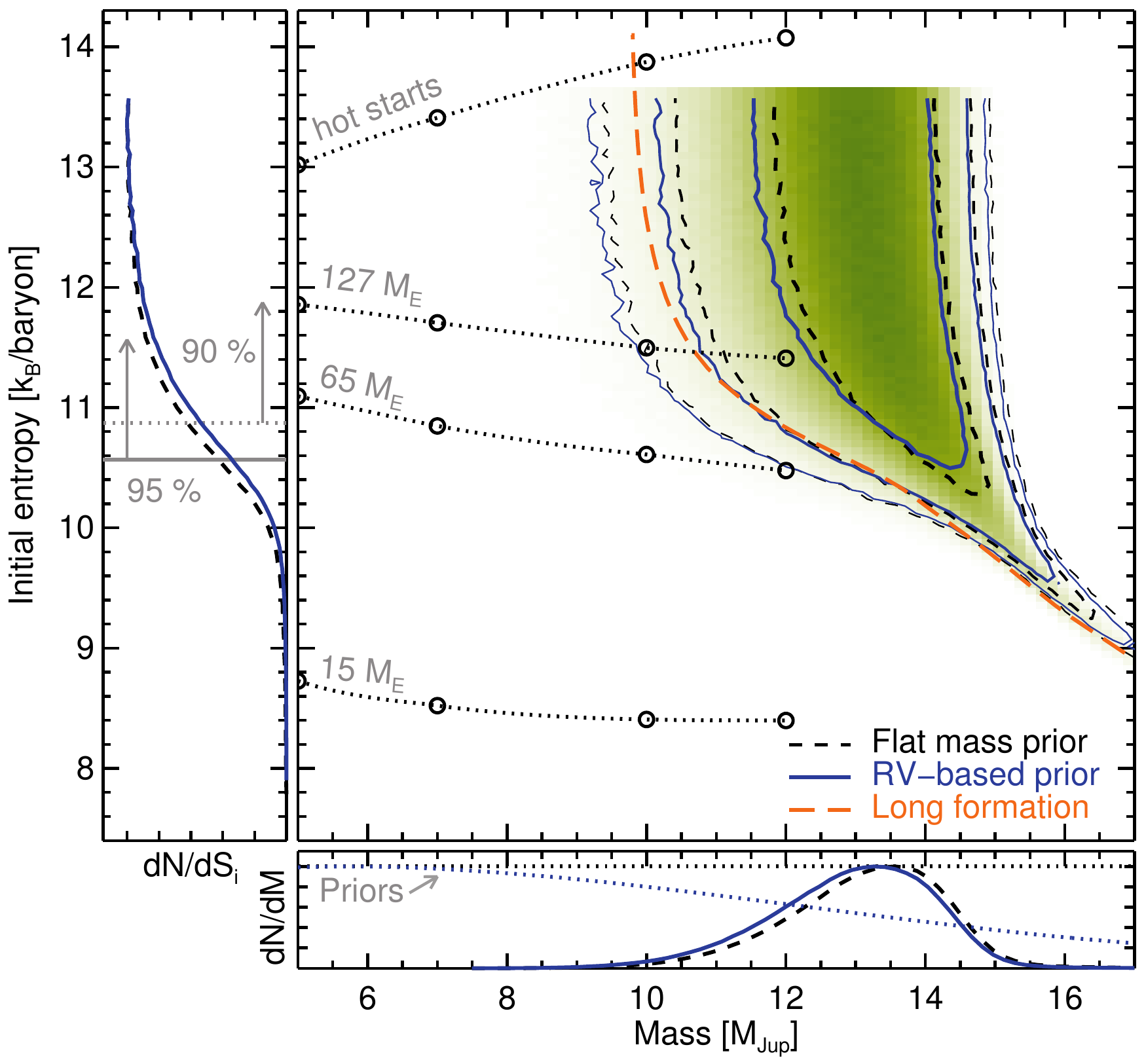}
\caption{Joint probability contours (68.3, 95, and 99\%) on the mass and post-formation entropy of $\beta$ Pic b from $\log(L/L_{\odot}) = -3.90\pm0.07$ and an age of $21\pm4$~Myr using a flat prior on $S_i$ and a prior flat in mass (dashed black curves) or given by the radial-velocity and astrometry constraints for a linear $K$ prior (solid dark-blue curves). Using the logarithmic-$K$ prior of Fig.~1 gives nearly identical results. The long-dashed orange line shows the combinations matching $\log (L/L_{\odot}) = -3.90$ at 12~Myr, i.e.\ allowing for an extreme 9 Myr formation delay. The open circles indicate the cold-start post-formation entropies for different final core masses (labeled) or for hot starts \citep[][App.~B of MC14]{2013A&A...558A.113M}. The bottom panel displays the mass priors (dotted lines) and the marginalized posterior distributions (black and blue lines), whereas the side panel shows the marginalized $S_i$ posterior and the non-flat mass prior's 90- and 95-\%\ lower limits (up to $S_i\sim14$).}
\label{Fig:Fig4}
\end{figure}

To derive quantitative constraints on the initial entropy $S_i$ of $\beta$ Pic~b, we used the method of MC14 and performed an MCMC in mass and $S_i$ using  their evolutionary models up to masses of 17 $\mathrm{M_{Jup}}$. The models have gray atmospheres, include deuterium burning (Marleau \&\ Cumming, in prep.), and span in $S_i$ the extreme outcomes of any formation process.  Fig. \ref{Fig:Fig4} shows the allowed $M$ and $S_i$ combinations that match the luminosity and age taking Gaussian errorbars into account.

%


\section{Discussion}
If the system is truly $21\pm4$~Myr old, Fig. \ref{Fig:Fig4}  shows that $\beta$ Pic b cannot have formed according to the classic \citep{2007ApJ...655..541M} parameters of core accretion, which include a supercritical accretion shock (coldest starts) and an initial planetesimal density leading to a 15-$M_{\oplus}$ core. An on average inefficient shock and/or a higher planetesimal density \citep{2013A&A...558A.113M}  must be invoked to lead to warmer starts. For a completely efficient accretion shock, the predicted core would need to be $\gtrsim65~M_{\oplus}$. These conclusions are nearly unchanged even assuming an extreme duration for the planet formation phase of 9~Myr (Fig.~\ref{Fig:Fig4}).

Moreover, we find a lower bound\footnote{An upper limit is given by the fact that the radius starts to diverge when $S$ increases above $\simeq16.5$. Varying the upper bound of the cumulative integral barely varies the quoted values.} on the post-formation entropy of $S_{i,\textrm{min}} = 10.5~k_{\rm B}\,\rm{baryon}^{-1}$ at the 95-\%\ level, which is $\approx2~k_{\rm B}\,\textrm{baryon}^{-1}$ warmer than the supercritical 15-$M_{\oplus}$ prediction. Finally, for masses within the 68.3-\%\ contour, the MC14 cooling curves predict $\beta$ Pic b to not be affected by deuterium flashes  \citep{2013ApJ...770..120B}, where the luminosity and $\mathrm{T_{eff}}$ of massive objects increase, possibly at very late times (MC14; Marleau \& Cumming, in prep.). However, because of differences in boundary conditions and nuclear rate details, and given the high precision of the luminosity measurement, using other cooling tracks can somewhat affect the mass constraints and the importance of deuterium burning in the cooling history of $\beta$~Pic~b.

%

\bibliographystyle{aa}
\bibliography{Bpic_GPI}

\begin{acknowledgements}
We thank the consortium who built the GPI instrument. We especially wish to thank C. Marois, D. Mouillet, C. Mordasini, J. Faherty and A. Triaud for fruitfull discussions. We are grateful to J. Gagn\'{e}, D. Lafreni\`{e}re, M. Liu, A. Schneider, K. Allers, J. Patience, E. Rice, N. Lodieu and E. Manjavacas for providing their spectra of young objects. We aknowledge P. Hauschildt and S. Witte for their work on the DRIFT-PHOENIX models. This research has benefitted from the SpeX Prism Spectral Libraries, maintained by A. Burgasser. MB, GC, AML, JR, HB, FA, and DH acknowledge financial support from the French National Research Agency (ANR) through project grant ANR10-BLANC0504-01, ANR-07-BLAN-0221, ANR-2010-JCJC-0504-01 and ANR-2010-JCJC-0501-01. ChH and DH highlights EU financial support under FP7 by starting grand. JLB PhD is funded by the LabEx Exploration Spatiale des Environnements Planétaires (ESEP) \# 2011-LABX-030.
\end{acknowledgements}

\Online
 \begin{appendix} 
 \section{Details on the orbital fit}
 \label{Appendix:Appmass}
 \subsection{Errors on the radial velocities}
The RV of $\beta$ Pictoris A measured within the same day are extremely variable because of the activity of the star. During the HARPS monitoring of the star, $\beta$ Pic A was either observed multiple times during a night to evaluate and average the stellar activity, or at a single time (Borgniet et al. in prep.). We averaged the data over one day to estimate a daily RV mean. To estimate the error on the RV corresponding to a night that properly account for the stellar noise, we assumed that the intrinsic RV variability is sinusoidal:  $A \times sin(\omega \times t+\phi)+C(t)$, with $A$ the amplitude of the variability, $\omega$ the angular frequency, $t$ the time, $\phi$ the phase, and $C$ an offset velocity. Radial velocity measurements performed over one single day can be regarded as successive values of a random variable following this law with random $t$. The resulting random RV has the following probability function:\\

\begin{equation}
                          p(RV) \times d(RV) = \dfrac{(1/\pi) \times 1}{\sqrt{A^{2}-(RV-C)^{2}}} \times d(RV).
\end{equation}

The variance of this law is $A^{2}/2$. Taking the arithmetic mean of $n$ independent measurements over one night gives an estimate of the offset velocity $C$  with $A/\sqrt{2n}$ as error. We now need to estimate $A$. We assume that $C$ varies with time, but $A$ does not. For a given day with the highest number of measurements N, the statistical variance $SN$ of these data is calculated. An accurate, unbiased estimator of $A^{2}$ is $2 \times (N/(N-1)) \times SN$, so that for any other day with $n$ measurements the error can be estimated to be $\sqrt{s^{2}+A^{2} \times N/(N-1)/n}$, where $s$ is the mean of the given HARPS RV measurment errors of the day. This way, errors are reduced for a day with many measurements and kept large for days with one or two measurements.\\
 
 \subsection{Choices of the priors}
Priors on the orbital parameters are identical to those used in \cite{2012A&A...542A..41C} when only the system astrometry is accounted for in the orbital fit. Changes to them appear to have little influence on the posterior distributions of orbital parameters.  In contrast, this is not the case for the mass determination because of the weak constraint provided by the radial velocity data. The most straightforward prior we can assume for the amplitude of the RV curve of $\beta$ Pic A  $K$ is linear, but a logarithmic prior (linear in $\ln K$) is also worth considering because $K$ is proportional to $P^{-1/3}$ (where $P$ is the orbital period), and a logarithmic prior for $P$ was already assumed. Figure \ref{Fig:Fig1}  shows the posterior mass determination for both priors. Because of the activity of the star, the data are compatible with planet masses down to virtually 0. But a lower cutoff at 2 $\mathrm{M_{Jup}}$  was assumed to remain compatible with the observed luminosity of the planet. The linear prior nevertheless appears to favor larger masses than the logarithmic prior. Then the major difference resides in the shape of the posterior distribution. The linear prior exhibits a clear peak around 6 $\mathrm{M_{Jup}}$. This difference illustrates the difficulty in deriving a relevant fit of the mass of $\beta$ Pic b. Obviously, the RV data are too noisy to allow a clear determination, but i) a conservative upper limit is confirmed, and ii) the peak around 6 $\mathrm{M_{Jup}}$ needs to be confirmed with future data.

 \section{Samples of comparison spectra}
 \label{Appendix:AppA}
For the purpose of the empirical analysis, we used four samples of spectra of ultracool MLT dwarfs found in the literature. The SpecXPrism library\footnote{http://pono.ucsd.edu/$\sim$adam/browndwarfs/spexprism} represents sample 1. Sample 2 is composed of spectra of M and L dwarfs with features indicative of low surface gravity \citep{2013ApJ...772...79A, 2014A&A...564A..55M, 2013ApJ...777L..20L, 2014AJ....147...34S}. The third sample is made of spectra of  members of 1-150 Myr old young moving groups and clusters \citep{2008MNRAS.383.1385L, 2010ApJS..186...63R,2014A&A...562A.127B, 2014arXiv1403.3120G}. The fourth sample is composed of spectra of young  MLT companions \citep{2010A&A...517A..76P, 2010ApJ...719..497L, 2011ApJ...729..139W, 2010A&A...512A..52B, 2014A&A...562A.127B}.

 \section{Description of the LESIA model grids}
 \label{Appendix:AppB}
\cite{2014IAUS..299..277B} developed a radiative-convective equilibrium model for young giant
exoplanets in the context of direct imaging. The input parameters are the planet
surface gravity (log g), effective temperature ($T_{\rm eff}$), and elemental
composition. Under the additional assumption of thermochemical
equilibrium, the model predicts the equilibrium-temperature profile and
mixing-ratio profiles of the most important gases. Opacity sources
include the H$_{2}$-He collision-induced absorption and molecular lines from
H$_{2}$O, CO, CH$_{4}$, NH$_{3}$, VO, TiO, Na, and K. Line opacity is modeled using
k-correlated coefficients pre-calculated over a fixed
pressure-temperature grid. Absorption by iron and silicate cloud
particles is added above the expected condensation levels with a fixed
scale height and a given optical depth at some reference wavelength.
To study $\beta$ Pic b, we built five grids of models with $T_{\rm eff}$ between 700 and 2100 K (100 K increments), log g between 2.1 and 5.5 dex (0.1 dex increments), and solar system abundances \citep{2010ppc..conf..379L}.   One model grid was created without clouds (hereafter set I). We added three grids with cloud particles located between condensation level and a 100 times lower pressure, with a particle radius of 30 $\mu$m ($\tau$=0.1, 1, 3; hereafter set II, III, IV), a scale height equal to the gas scale height, and optical depths ($\mathrm{\tau_{cloud}}$) of 1$\tau$ and 0.15$\tau$ at 1.2 $\mu$m for Fe and Mg$_{2}$SiO$_{4}$, respectively (assuming the same column density for both clouds). We used an additional grid (hereafter set V) with a particle radius of 3 $\mu$m and $\mathrm{\tau_{cloud}}$ of 1 and 0.018. The grid properties are summarized in Table \ref{Table:TableAppB}.

\begin{center}
\begin{table}[htbp]
\begin{minipage}{\linewidth}
\caption{Properties of the LESIA atmospheric model grids}
\label{Table:TableAppB}
\centering
\renewcommand{\footnoterule}{}  
\begin{tabular}{cccc}
\hline \hline
Model & $\tau_{Fe} $     & $\tau_{Mg_{2} SiO_{4}}$ & Particule radius \\
           &  (1.2 $\mu$m)  &  (1.2 $\mu$m)   &  ($\mu$m)    \\
\hline
I  & 0 & 0 & 0\\
II & 0.1&0.015 & 30\\
III & 1 &0.15 & 30 \\
IV & 3 &0.45 & 30 \\
V & 1 & 0.018 & 3 \\
           \hline
\end{tabular}
\end{minipage}
\end{table}
\end{center}

 \section{Fit of the spectral energy distribution}
 \label{Appendix:AppC}

 The planet SED was built from the $Y_{s}$ and $CH_{4S,1\%}$ band photometry reported reported in \cite{2014ApJ...786...32M}, $J$,$H$, $L'$ and $M'$ band photometry \cite{2013A&A...555A.107B}, $K_{s}$-band photometry from \cite{2013ApJ...776...15C}, and $NB_{4.04}$ band magnitude from \cite{2010ApJ...722L..49Q}. The SED and spectral-fitting procedures are described in \cite{2013A&A...555A.107B} and \cite{2014A&A...562A.127B}, respectively. 

\begin{table*}
\caption{Same as Figure \ref{atmopar}, but for the spectral energy distribution fit of $\beta$ Pic b. The semi-empirical radius $\mathrm{\mathrm{R_{S.E.}}}$ derived from $\mathrm{T_{eff}}$ and the bolometric luminosity can be compared with the radius inferred from the synthetic spectral fitting ($R$).}
\label{atmoparSED}
\centering
\renewcommand{\footnoterule}{}  
\begin{tabular}{lllllll}
\hline \hline 
Model		 	   &    $\mathrm{T_{eff}}$		&	 $\mathrm{log\:g}$	& R		& $\mathrm{\chi^{2}_{red}}$  & $\mathrm{\mathrm{M_{S.E.}}}$	  &  $\mathrm{\mathrm{R_{S.E.}}}$ \\   			
								&				(K)					&		$\mathrm{(cm.s^{-2}}$)	&	($\mathrm{R_{Jup}}$) & &	($\mathrm{M_{Jup}}$)&	($\mathrm{R_{Jup}}$)  	\\	
\hline
Settl 10			&		1600						&	4.0			&	1.57		&	0.82	  &  8$^{+21}_{-6}$ & $1.4\pm0.1$ \\
Settl 13-M/H=-0.5	 &		1800						&	3.5			&	1.24		&	1.34  &  2$^{+4}_{-2}$ & 	$1.1\pm0.1$ \\
Settl 13-M/H=0.0	& 		1800						&	4.0			&	1.26		&	1.42	& 5$^{+13}_{-4}$ &  $1.1\pm0.1$ \\
Settl 13-M/H=+0.5	& 		1700						&	5.0			&	1.61		&	0.64	&  $64^{+157}_{-46}$ &  $1.3\pm0.1$ \\
DP-M/H=-0.5            &   1700              &  4.0               &  1.43      &  0.38  &   6$^{+16}_{-5}$ &  $1.3\pm0.1$ \\
DP-M/H=0.0            & 1800                   &  4.5             &  1.27      &  0.52    & 16$^{+39}_{-12}$ & $1.1\pm0.1$\\
DP-M/H=+0.5            &  1800                &  4.5             &   1.34     &  0.66   & 16$^{+39}_{-12}$ & $1.1\pm0.1$\\
LESIA - I           &   \textit{1600}                &  \textit{2.1}            &    \textit{1.58}     &   \textit{2.38}  & $0.1\pm0.1$  & $1.4\pm0.1$ \\
LESIA - II          &   \textit{1400}                 &  \textit{5.5}            &   \textit{1.95}      &   \textit{0.66}  & 441$^{+195}_{-129}$  &  $1.9\pm0.1$ \\ 
LESIA - III         &   1500                 &  3.8            &    1.76     &   0.50   & 7$^{+3}_{-2}$  &   $1.6\pm0.1$\\
LESIA - IV        &    1500                 & 3.2            &    1.78     &  0.60   & 1.7$^{+0.7}_{-0.5}$ & $1.6\pm0.1$ \\
LESIA - V         &    1600                 &  4.1             &    1.56    &   0.72  & 10 $^{+4}_{-3}$ & $1.4\pm0.1$ \\
\hline
\end{tabular}
\end{table*}

\begin{figure}[htbp]
\centering
\includegraphics[width=8.6cm]{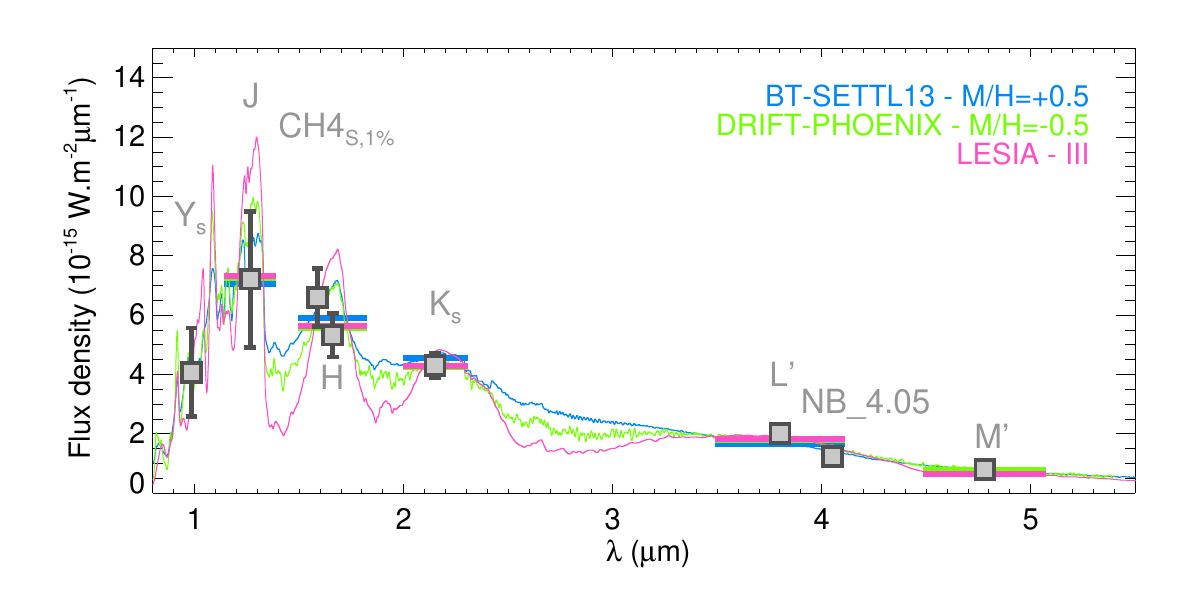} \\
\caption{Comparison of $\beta$ Pic b SED to best-fitting synthetic spectra (solid line) and fluxes (horizontal lines) from the BT-SETTL,  DRIFT-PHOENIX, and LESIA atmospheric models grids.}
\label{Fig:SEDfit}
\end{figure}

 \end{appendix}


\end{document}